\documentclass[aip,preprint,graphicx]{revtex4-1}
\usepackage{graphicx}

\draft 

\begin{document}

\title{Effect of Self-Magnetic Fields on the Nonlinear Dynamics of
Relativistic Electron Beam with Virtual Cathode}

\author{A.E. Hramov}
\email[Electronic mail: ]{hramovae@gmail.com} \affiliation{Saratov
State University, Astrakhanskaja 83, Saratov 410012, Russia\\
Saratov State Technical University, Politechnicheskaja 77, Saratov
410028, Russia}

\author{S.A. Kurkin}
\email[Electronic mail: ]{KurkinSA@gmail.com} \affiliation{Saratov
State University, Astrakhanskaja 83, Saratov 410012, Russia}

\author{A.A. Koronovskii}
\email[Electronic mail: ]{alexey.koronovskii@gmail.com}
\affiliation{Saratov
State University, Astrakhanskaja 83, Saratov 410012, Russia\\
Saratov State Technical University, Politechnicheskaja 77, Saratov
410028, Russia}

\author{A.E. Filatova}
\email[Electronic mail: ]{anefila@gmail.com}
\affiliation{Saratov State Technical University, Politechnicheskaja 77, Saratov
410028, Russia}

\date{\today}

\begin{abstract}
The report is devoted to the results of the numerical study of the
virtual cathode formation conditions in the relativistic electron
beam under the influence of the self-magnetic and external axial
magnetic fields. The azimuthal instability of the relativistic
electron beam leading to the formation of the vortex electron
structure in the system was found out. This instability is
determined by the influence of the self-magnetic fields of the
relativistic electron beam and it leads to the decrease of the
critical value of the electron beam current (current when the
non-stationary virtual cathode is formed in the drift space). The
typical dependencies of the critical current on the external uniform
magnetic field value were discovered. The effect of the beam
thickness on the virtual cathode formation conditions was also
analyzed.
\end{abstract}


\maketitle

\section{Introduction}
\label{Intro} The analysis of nonlinear microwave oscillations and
complex structure formation mechanisms in spatially extended systems
with intensive beams of charged particles in the regimes of the virtual
cathode (VC) formation attracts great attention of scientific
community \cite{Sullivan:1987_VCO_Book, Alyokhin:1994,
Dubinov:2002_ReviewEngl, Kalinin:2005_PhysPlazma_ENG, Benford:2007_book, biswas:063104,
filatov:033106, ender:033502, Singh:2011_VC}. It is well known~\cite{Dubinov:2002_ReviewEngl,
Alyokhin:1994, Kalinin:2005_PhysPlazma_ENG, Hramov:2010_MPFS} that the systems with
VC are characterized by the complex dynamics and can demonstrate a
wide range of nonlinear phenomena, including dynamical chaos and
pattern formation \cite{Trubetskov:1996_CHAOS,
Anfinogentov:1998_VCO_ENGL, Koronovskii:2002_PierceEnglish, hramov:2011_IJE}.
Microwave generators using electron beams with VC
(vircators\cite{Mahaffey:1977_First_VC, Sullivan:1987_VCO_Book}) are
perspective devices of high-power microwave electronics for the
generation of the impulses of wide-band microwave radiation
due to its high output microwave power, a simple construction
(particularly vircators can operate without external focusing
magnetic field), a possibility of a simple frequency tuning and
regime switching \cite{Sze:1986_VC, Dubinov:2002_ReviewEngl,
Burkhart:1985_VCO, Hoeberling:1992_AdvanceVCO, Benford:2007_book}.
All these circumstances increase the
fundamental and applied importance of studies of the
nonlinear dynamics of the electron beams with VC.

The oscillating VC is known to appear in an electron beam when the
beam current exceeds a certain critical value, $I_{cr}$ (space
charge limiting (SCL) current) \cite{Bogdankevich:1971_Engl,
Sullivan:1987_VCO_Book}, and the beam space charge is strong enough
in order to form a potential barrier (VC) which reflects the
electrons back to the injection plane. The mechanisms of the VC
formation have been investigated in detail in case of the
one-dimensional (1D) electron beam motion (fully magnetized
beam)~\cite{Jiang1995_Mech_VC, Anfinogentov:1998_VCO_ENGL,
Hramov:1999_ChaosVCO, Egorov_1D:2007, filatov:033106}, with the critical
beam current value being analytically defined for this case in
Ref.~\cite{Bogdankevich:1971_Engl}.

However, the use of 1D-theory to study the vircator systems is
inefficient in many cases because it ignores a lot of
important factors in the VC behavior and does not agree often with
the experimental results. So, the development and use of 2D and 3D
models for the analysis of the dynamics of electron beam with VC
have attracted the great interest of scientific community
recently~\cite{Lindsay:2002_VCO, Chen:2004_VCO_Coaxial, Hramov:2008_Critical_Current,
Hramov:2010_MPFS}. The important problem in this field is the
analysis of the VC formation conditions and nonlinear dynamics of
electron structures in relativistic electron beams (REB) in the
presence of finite external magnetic field or even without external
magnetic field which focuses the REB. In particular,
the studies of REBs with overcritical currents are necessary for the
analysis of contemporary high-power devices with VC -- relativistic
vircators \cite{Benford:2007_book} and ion acceleration systems
\cite{Magda:1997, Dubinov:2002_ReviewEngl}.

Analyzing REBs, it is necessary to take into account effects being
insignificant for weakly relativistic beams, in particular, the
influence of the self-magnetic field of the REB that effects
considerably on the system dynamics in case of ultra-relativistic
electron beams. Therefore, the 3D fully electromagnetic
self-consistent model of REB dynamics is required for accurate and
correct analysis of the VC formation in this case.

The present report deals
with the 3D numerical electromagnetic study of the VC nonlinear
dynamics of the annular REB in the presence of an external finite
uniform  axial magnetic field. The structure of this paper is the following.
Section~\ref{sct:model} contains the brief formalism describing a 3D
mathematical model for the nonlinear interaction simulation of
electron beam with overcritical current and electromagnetic fields.
Section~\ref{sct:nonlindynamics} deals with the nonlinear dynamics
of REB with overcritical current. We analyze the influence of the
external magnetic field on the VC formation conditions and structure
formation in REB and discuss the azimuthal REB instability which
leads to the formation of a vortex electron structure.
In Section~\ref{sct:thickness} the influence of the beam thickness on the VC formation
conditions is analyzed. The conclusions of this paper
are summarized in Section~\ref{sct:concl}.

\section{General formalism}
\label{sct:model} The model under study consists of finite-length
cylindrical waveguide region (electron beam drift chamber) with
length $L$, radius $R$ and grid electrodes at both ends. An
axially-symmetrical monoenergetic annular electron beam with the
current $I$, electron energy $W_e$, radius $R_b$ and thickness $d$
is injected through the left (entrance) electrode. Electrons can
leave the waveguide region by escaping through the right (exit) grid
or by touching the side wall of the drift chamber. In the present
work the values of geometric parameters were chosen following:
$L=40$\,mm, $R=10$\,mm, $R_b=5$\,mm, $d=1.5$\,mm (except for
the Section~\ref{sct:thickness} where the influence of the beam
thickness on the VC formation conditions is analyzed).

The external uniform magnetic field with induction $B_z=B_0$ is
applied along the longitudinal axis of waveguide. The electron beam
source is supposed to be magnetically unshielded in the considered
model. This assumption means that the
external magnetic field in the drift tube is equal to the magnetic
field in the electron source region, therefore, the electron beam
doesn't acquire azimuthal velocity components at the injection plane
(in accordance with Busch's theorem~\cite{Tsimring:2007_Beams}).
Such magnetic field distribution is typical for many devices of the
high-power electronics, particularly for magnetically isolated
diodes that forms high-current REBs~\cite{Tsimring:2007_Beams}.

Time-dependent fully 3D electromagnetic model of REB dynamics based
on solving the self-consistent set of Maxwell equations and
equations of charged particles motion accompanied by corresponding
initial and boundary conditions (particle-in-cell
method)~\cite{Birdsall:1985_PlasmaBook,
Anderson:1999_EM_Simulations} is used in the present paper.
The main equations are written as:
\begin{equation}\label{maxvell_eq1}
 {\rm rot}\,{\bf  E}=-\frac{1}{c}\frac{{\partial\bf  H}}{\partial t},\qquad
 {\rm rot}\,{\bf  H}=\frac{1}{c}\frac{{\partial\bf  E}}{\partial t}+\frac{4\pi}{c}{\bf  j},
\end{equation}
\begin{equation}\label{Particles}
\frac{{d\bf p}_i}{dt}={\bf E}_i+[{\bf p}_i, {\bf B}_i]/\gamma_i,\quad
\frac{{d\bf  r}_i}{dt}={\bf  p}_i/{\gamma_i},\quad i=1\dots N,
\end{equation}
where $\bf E$ and $\bf H$ are the electric and magnetic intensities,
$\rho$ and $\bf j$ are the charge and current densities, $\bf r$,
$\bf p$, $\gamma=\left(1-(v/c)^2\right)^{-1/2}$, $\bf v$ are the
radius vector, impulse, relativistic factor and velocity of the
charged particles, correspondingly. The subscript $i$ denotes the
number of particle and $N$ is the full number of particles used to
simulate the charged particles beam.

The numerical simulation 3D scheme is based on 2.5D model
developed in our previous work \cite{Egorov:2006_2.5DVCO_Engl}. The
equations of charged particles motion (\ref{Particles}) are used for
electron beam simulation and solved numerically by means of
B\'oris algorithm \cite{Boris:1969}. The longitudinal  $v_z$, radial
$v_r$ and azimuthal $v_\theta$ velocity components are calculated
with the help of this algorithm on each time step.

The electromagnetic fields in the drift chamber of REB are obtained
by means of the numerical solution of the Maxwell's equations
(\ref{maxvell_eq1}) in cylindrical geometry on the shifted
spatio-temporal meshes with constant spatial longitudinal, $\Delta
z$, radial, $\Delta r$, and azimuthal, $\Delta\theta$, steps and
time step, $\Delta t$ \cite{Birdsall:1985_PlasmaBook,
Anderson:1999_EM_Simulations}. The values of steps of
spatio-temporal meshes is picked out from the
Courant-Friedrichs-Lewy condition
\cite{Rouch:1976_FluidNumericalBook, Birdsall:1985_PlasmaBook}. The
every field component is calculated on the own mesh (see
\cite{Birdsall:1985_PlasmaBook, Anderson:1999_EM_Simulations}). The
space charge and current densities on the meshes are calculated
using a bilinear weighing procedure~\cite{Birdsall:1985_PlasmaBook}.
To model the electromagnetic power output we use the approach~\cite{Egorov:2006_2.5DVCO_Engl}
based on the filling the section of electrodynamical system
($L<z<1.2L$) with the conducting medium
with the conductance $\sigma$.

\section{Nonlinear dynamics of relativistic electron beam with overcritical
current} \label{sct:nonlindynamics}

\subsection{Virtual cathode formation conditions}
\label{sct:VCO_form}

Let us consider the results of numerical simulation of VC
formation features in the annular REB in the presence of the external
magnetic field. Here we study the influence of the magnetic field on the
critical current of REB (i.e., VC formation conditions) as well as on the
nonlinear dynamics of the electron structures in the REB. External
magnetic field, $B_0$, varies within the range $[0, 40]$\,kGs.

\begin{figure}[h]
  \includegraphics[width=8.2cm]{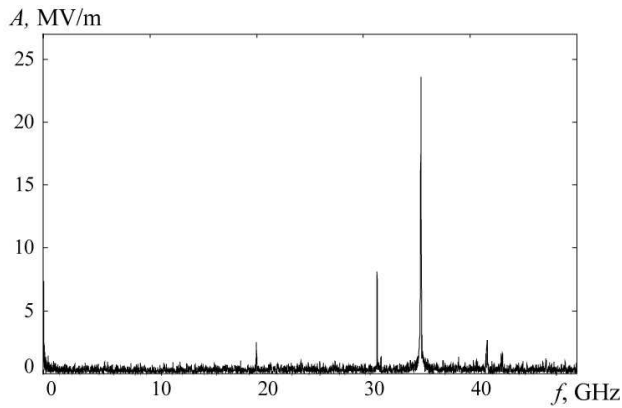}
    \caption{Amplitude spectrum of oscillations of self-electric field longitudinal component $E_z$ at the VC region for $I=17$\,kA, $B_0=0$}
    \label{Pic:Spectr}
\end{figure}

VC in the electron beam with overcritical current
is characterized by the complex non-stationary spatial-temporal oscillations
\cite{Granatstein:1987_Book, Alyokhin:1994} that leads to the excitation of
vircator's electrodynamic system, and, consequently, to the generation
of high-power microwave radiation in such system.
Fig.~\ref{Pic:Spectr} demonstrates the numerically obtained typical
amplitude spectrum of oscillations of self-electric field
longitudinal component at the VC region in the considered vircator
system. The carried out electrodynamic analysis has shown that
frequencies of spectral components in Fig.~\ref{Pic:Spectr} are
determined by the corresponding eigenmodes of the
electrodynamic system formed by the drift chamber walls. REB critical current is
supposed to correspond to such value of the beam current when the spectral components appear at the output spectrum
of vircator system and electrons start to reflect back to the
injection plane $z=0$.

\begin{figure}[h]
  \includegraphics[width=8.2cm]{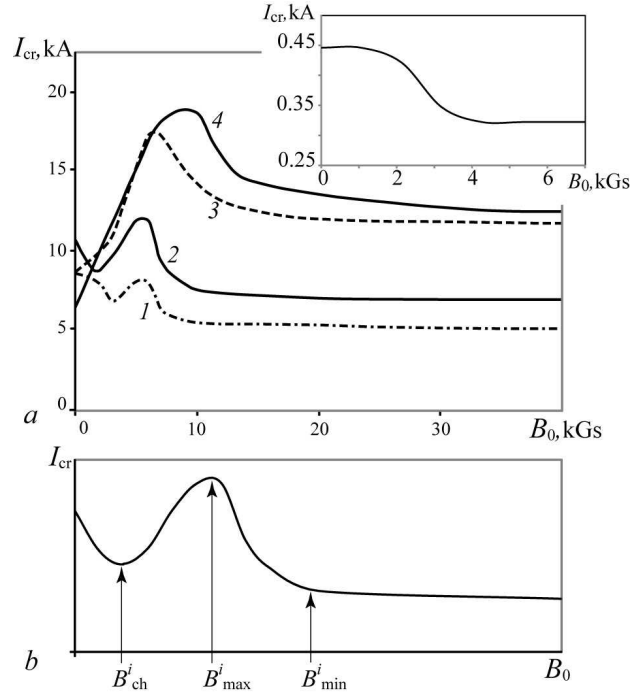}
\caption{(\textit{a}) Dependencies of the REB critical current on the
external magnetic field induction value for the following electron
beam initial energy values $W_e$: curve~\emph{1} corresponds to
480\,keV, \emph{2} -- 600\,keV, \emph{3} -- 850\,keV and \emph{4} --
1\,MeV. The insert panel demonstrates the dependency of the beam
critical current on the magnetic field $B_0$ in the weakly relativistic case
(beam energy $W_e=79$\,keV) \cite{Kurkin:2009_PPR,
Hramov:2008_Critical_Current}. (\textit{b}) Illustration of the typical
REB curve $I_{cr}(B_0)$ with denoted character values of the
external magnetic field, where $B_{ch}^i$ is the value of the magnetic
field when the critical beam current reaches the first minimum;
$B_{max}^i$ is the magnetic field when the critical beam current reaches the
local maximum and  $B_{min}^i$ is the magnetic field when the curve
$I_{cr}(B_0)$ saturates}
    \label{Pic:CrB}
\end{figure}

Fig.~\ref{Pic:CrB}a shows the dependencies of REB critical current,
$I_{cr}$, on the external magnetic field value, $B_0$, for different
beam energies, $W_e$. The insert panel in Fig.~\ref{Pic:CrB}a
demonstrates the corresponding critical current values for weakly
relativistic electron beam with the initial electron energy
$W_e=79$\,keV (see Ref.~\cite{Kurkin:2009_PPR,
Hramov:2008_Critical_Current}). Analyzing this Figure, one can see
that the curves $I_{cr}(B_0)$ for REB have characteristic feature
in comparison with the weakly relativistic case when the critical current
decreases monotonously approaching the minimal constant value with
the growth of $B_0$.

The typical shape of REB critical current dependency on
external magnetic field value is shown in Fig.~\ref{Pic:CrB}b. As one can see from
Fig.~\ref{Pic:CrB}, the REB critical current
increases with the growth of induction, $B_0$, within the range ${B<B_{max}^i}$.
With further growth
of magnetic field for $B_0>B_{max}^i$ the REB critical current
decreases and for $B_0\sim B_{min}^i$ the saturation at the constant
level is observed. Electron beam appears to be fully magnetized in the
case of the strong external magnetic field $B_0>B_{min}^i$ and moves
generally one-dimensionally. Bogdankevich and Ruhadze were shown
\cite{Bogdankevich:1971_Engl}
that the analytical expression for the critical current value in the
case of 1D motion of electron beam has the following form:
\begin{equation}\label{I_0_B=INFTY}
I_{SCL}=\frac{c^3}{\eta}\frac{\left(\gamma_0^{2/3}-1\right)^{3/2}}{d/R_b+2\ln
(R/R_b)},~~~ \gamma_0=\frac1{\sqrt{\left(1-(v_0/c)^2\right)}},
\end{equation}
where $R$ is the radius of cylindrical drift tube, $R_b$ è $d$ are
the radius and thickness of annular REB, $\gamma_0$ and $v_0$ are
the relativistic factor and velocity of REB electrons at injection
plane,  $c$ is the speed of light, and $\eta$ is the specific
electronic charge. Table~\ref{tab1} demonstrates the comparison of
REB critical current values obtained analytically using relation
(\ref{I_0_B=INFTY}) with the results of numerical simulation in the
case of the strongly magnetized beam ($B_0=40$~kG) for the different
values of the electron energy $W_e$. Analyzing Table~\ref{tab1}, one
can see that the numerical results for REB critical currents agree
accurately with the values obtained analytically, and, therefore, we
can conclude that the developed numerical 3D fully electromagnetic
model gives correct results. Note also, that the numerically
obtained critical current value exceeds slightly the analytical
results. It's a consequence of the 2D dynamics of the electron beam
occurring in the system even in the case of the strong external
magnetic field, e.g. the cyclotron rotation of electrons, pulsations
of the beam boundary et al. These 2D effects reduce the charge
density in the VC area and, consequently, enlarge weakly the
critical current value in comparison with 1D estimation
(\ref{I_0_B=INFTY}) which doesn't take into account these effects.

It should be noted that the effects of the decrease of the space charge
limiting current in comparison with 1D electrostatic case
(\ref{I_0_B=INFTY}) \cite{Bogdankevich:1971_Engl} due to the
electromagnetic transients in the injected currents defined by the
inductive voltage, $L{dI}/{dt}$, and the power loss in the drift
chamber through the open boundary have been reported earlier~\cite{700866, biswas:073101}.
However, these phenomena do not play the significant role for the
system under consideration. Actually, the inductive transient effect
is pronounced only for the short drift chambers (${W}/{L}>1$, where $W=2R_b$ is the
beam diameter, and $L$ is the drift chamber length) \cite{700866},
while for the considered system ${W}/{L}=0.25$. The power loss
is also negligible in our simulations, so its influence on the critical
current value is insignificant in the considered case
\cite{biswas:073101}.

Thereby, the carried out analysis has shown (Fig.~\ref{Pic:CrB}a)
that REB critical current curves have local maxima at the
certain external magnetic field value $B_0=B_{max}^i$ dependent on the
beam energy $W_e$. Such behavior is not observed in the weakly
relativistic electron beams and develops with the electron beam
energy growth.

\begin{table}
\caption{\label{tab1} Critical current values obtained analytically
using relation (\ref{I_0_B=INFTY}) ($I_{cr}^{an}$, third column) and numerically
($I_{cr}^{num}$, fourth column) for the different values of the beam energy ($W_e$, first column);
second column shows corresponding values of $v_0/c$; $B_0=40$~kG.}
\begin{ruledtabular}
\begin{tabular}{cccc}
$W_e$\,keV,&$v_0/c$&$I_{cr}^{an}$\,kA\footnote{Analytical results \cite{Bogdankevich:1971_Engl}}&$I_{cr}^{num}$\,kA\footnote{Numerical results}\\
\hline
480 & 0.8571 & 4.82 & 4.93\\
600 & 0.8882 & 6.45 & 6.61\\
850 & 0.9270 & 10.01 & 10.12\\
1000 & 0.9412 & 12.11 & 12.23\\
\end{tabular}
\end{ruledtabular}
\end{table}

\subsection{Structure formation in REB with overcritical current}
\label{sct:PhysProc}

The observed behavior of the REB critical current $I_{cr}(B_0)$, (see Fig.~\ref{Pic:CrB}a) is determined by the influence
of the self-magnetic fields of REB on the space charge dynamics and
electron structures formation. Therefore, this effect is more pronounced in the case of the weak external magnetic field and high energy
of the injected electron beam $W_e>600$\,keV, when the influence of the
self-magnetic fields of REB is stronger. Let us consider the effect
of electron structure formation in the presence of the REB
self-magnetic field in detail.

\begin{figure}[h]
  \includegraphics[width=8.2cm]{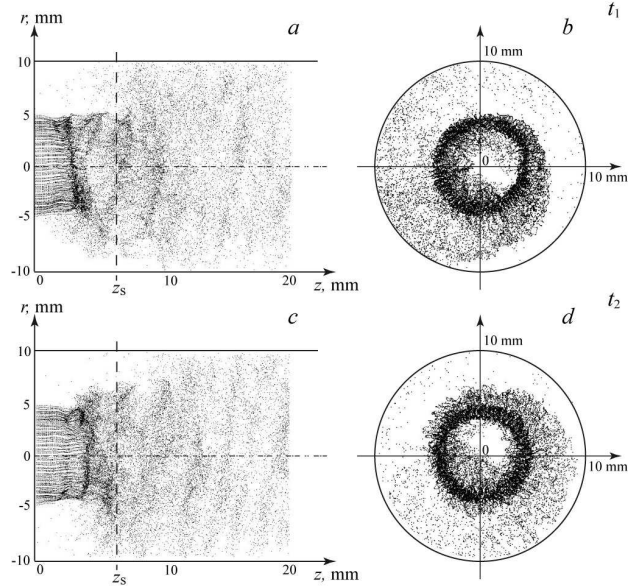}
    \caption{Projections of the instantaneous positions of the electron beam charged particles at the plane $(z,r)$ (left figures) and plane $(r,\theta)$ for $z_s=6$\,mm (right figures) at the consecutive time points $t_1$ and $t_2$ ($t_2-t_1=0.1$\,ns) for $B_0=3$\,kGs, $I=7.5$\,kA; $W_e=480$\,keV. Only particles behind the projection plane ($z>z_s$) are shown in Figures. The vertical dashed line with coordinate $z_s$ in figures \textit{a} and \textit{c} denotes the projection plane for figures \textit{b} and \textit{d}) }
    \label{Pic:REB1}
\end{figure}

\begin{figure}[h]
  \includegraphics[width=8.2cm]{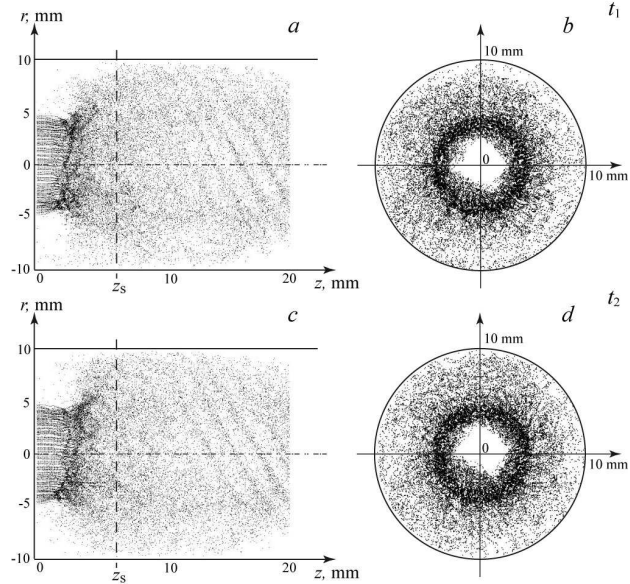}
    \caption{Similar to Fig.~\ref{Pic:REB1} but for the following parameters: $B_0=5$\,kGs, $I=9$\,kA; $W_e=480$\,keV; $t_2-t_1=0.2$\,ns}
    \label{Pic:REB2}
\end{figure}

\begin{figure}[h]
  \includegraphics[width=8.2cm]{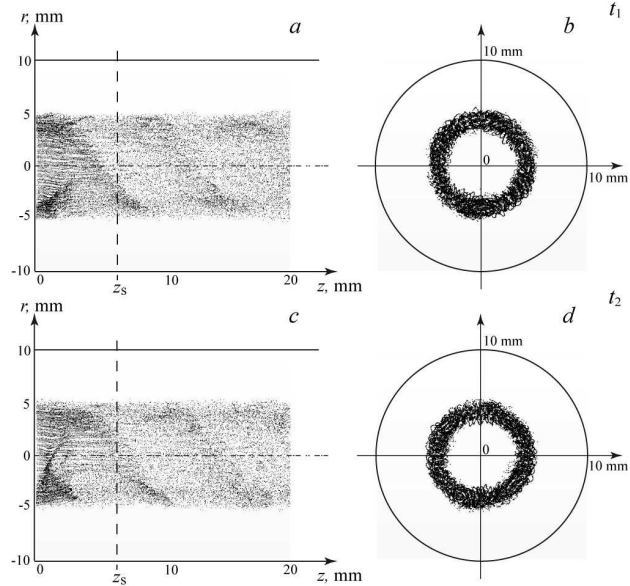}
    \caption{Similar to Fig.~\ref{Pic:REB1} but for the following parameters: $B_0=30$\,kGs, $I=10$\,kA; $W_e=480$\,keV; $t_2-t_1=0.2$\,ns}
    \label{Pic:REB3}
\end{figure}

Figures~\ref{Pic:REB1}--\ref{Pic:REB3} show the typical phase
portraits of the electron beam which are presented by the projections of the
instantaneous positions (black dots in figures) of charged particles
of the beam at the longitudinal $(z,r)$ and transverse planes
$(r,\theta)$ for the beam currents being close to the critical values $I_{cr}(B_0)$
($I\gtrsim I_{cr}(B_0)$) and the different characteristic values of the
external magnetic field, $B_0$. Fig.~\ref{Pic:REB1} corresponds to
the case $B_0<B_{max}^1$ when the external magnetic field value is less
than magnetic field $B_{max}^1$ corresponding to the local maximum on
curve \textit{1} in Fig.~\ref{Pic:CrB}a, Fig.~\ref{Pic:REB2} --- to
the case of $B_0\sim B_{max}^1$ and Fig.~\ref{Pic:REB3} --- to the
case when the magnetic field corresponds to the range of saturation of
$I_{cr}(B_0)$-curve on the constant level (i.e., $B_0\gg B_{min}^1$).

The space charge dynamics of the ultra-relativistic electron beam at
the VC region differs considerably in comparison with the weakly
relativistic case for the weak external magnetic field $0\leq B_0\leq
B_{max}^i$ (see Fig.~\ref{Pic:CrB}b). This major difference is determined by
the effect of the instability of
REB found for the wide range of beam currents in the case when
$0\leq B_0\leq B_{max}^i$. The observed instability, the so-called
azimuthal instability of REB, leads to the axial symmetry loss of
the electron beam dynamics.

This azimuthal instability of initially axially symmetrical beam in
cylindrical drift chamber facilitates the formation of the vortex
electron structure in the drift space. This instability is caused by the
influence of the self-magnetic fields of the REB (see
Fig.~\ref{Pic:REB1}). Actually, there is the intensive transverse beam
current in the system as a result of the beam divergence due to
Coulomb's repulsion forces action in the presence of the weak
external magnetic field which does not restrain these forces. This
transverse current results in the appearance of the longitudinal
self-magnetic field, $B^s_z$, that causes the azimuthal Lorentz
force action on the electrons moving in the transverse direction
\cite{Davidson:1974}. Therefore, these electrons get the azimuthal
velocities and REB starts to rotate around symmetry axis of the
drift space, $r=0$. Centrifugal force acts on rotating electrons and,
consequently, the vortex electron structure is formed in the beam
that results in the strong azimuthal asymmetry of the REB. The
formed vortex structure rotates in the drift tube as can be clearly
seen by comparing the configurational portraits in
Fig.~\ref{Pic:REB1}\textit{a} and \textit{c} for two consecutive
time moments. The physical mechanism of the azimuthal
instability development has similar features with the well-known
convective instability of the electron beam in the finite external
magnetic field \cite{levy:1288, PhysRevLett.54.1167} that arises
due to the non-uniform distribution of the beam electrons density or
velocity along the radial direction. However, the azimuthal instability
in the considered case arises due to interaction of the beam with the
self-magnetic, rather than external magnetic field.

The azimuthal instability leads to the decrease of the REB critical
current due to the decrement of the longitudinal velocity, $v_z$, and the
increase of the velocity spread of electrons in the region of the
formation of the vortex electron structure. As a consequence, VC is
formed in the region of the vortex structure onset, where the REB space
charge density is maximal due to a large number of electrons with
low energy.

Note, generally the vortex structure may be formed in the system
without the VC onset. The azimuthal instability is developed in the
REB when its current is greater than a certain threshold value. So,
if the REB current exceeds this threshold value and it is less than the
critical current (when the VC is observed in the system) only the
vortex structure is formed in the system.

With the growth of the external magnetic field value the azimuthal
instability becomes to be suppressed by the focusing force of the external magnetic
field which is directed oppositely to the centrifugal force.
Fig.~\ref{Pic:REB2} corresponding to the case of $B_0\sim B_{max}^1$
(see Fig.~\ref{Pic:CrB}a) demonstrates clearly this effect.
Actually, the rotating vortex electron structure (see
Fig.~\ref{Pic:REB1}\textit{b} and \textit{d}) disappears in the
drift space and the electron beam fills the whole space without the essential
azimuthal inhomogeneity. The space charge density decreases and, as
a consequence, the critical current for the VC formation increases
in this case. So, the suppression of the REB azimuthal instability
in the applied external magnetic field leads to the growth of the REB
critical current with the increase of the external magnetic field $B_0$.
Note, for the ultra-relativistic electron beams with energy $W_e>0.7$\,MeV
this increase starts simultaneously with the growth of the magnetic
field $B_0>0$ (curves\,3 and 4 in Fig.~\ref{Pic:CrB}a), whereas for the beams
with the less energy ($W_e\approx(0.4\div0.7)$\,MeV) the increase of the REB
critical current starts for the values of the external magnetic
field $B_0\geq B_{ch}^i$ (see curves\,1 and 2 in
Fig.~\ref{Pic:CrB}a).

A further growth of the magnetic field ($B_0>B_{max}^i$) leads to the decrease
of the REB critical current value and its saturation at constant level
for $B_0\sim B_{min}^i$ similarly to the weakly relativistic
case~\cite{Hramov:2008_Critical_Current, Kurkin:2009_PPR,
Kurkin:2009_JTF_VCMagnetic_Engl}. Such behavior of $I_{cr}(B_0)$ is
determined by the mechanism of the charged particle transversal
dynamics limitation by means of the external magnetic field (see
Fig.~\ref{Pic:REB3}). As a consequence, the space charge density of
the REB also increases with the growth of the external magnetic field,
and the critical beam current decreases. When the value of the external
magnetic field induction equals to $B_{min}^i$ the REB transversal
dynamics appears to be completely suppressed. Fig.~\ref{Pic:REB3}
shows that the transversal beam dynamics is not observed in the system
in the case of the strong magnetic field $B_0\geq B_{min}^i$.

The mentioned difference in the curves behavior in
Fig.~\ref{Pic:CrB} (cf. curves\,1 and 2 with curves\,3 and 4) for
the case of weak external magnetic fields $0\leq B_0\leq B_{ch}^i$
is the consequence of the competition of two described above
physical mechanisms occurring in the system with the growth of the
external magnetic field. The first mechanism (the suppression of the
azimuthal instability of the REB) leads to the critical current
value increase. The second one, which is connected with the charged
particle transversal dynamics limitation in external focusing
magnetic field, conducts  on the contrary to the decrease of
critical current. For the electron beams with energy
$W_e\approx(0.4\div0.7)$\,MeV (see curves\,1 and 2 in
Fig.~\ref{Pic:CrB}a) the both mechanisms have an important role: the
second mechanism determines the behavior of the dependencies for
small values of external magnetic field $0\leq B_0\leq B_{ch}^i$,
and the first one begins to dominate for $B_{ch}^i\leq B_0\leq
B_{max}^i$. In the case of ultra-relativistic electron beams with
energy $W_e>0.7$\,MeV when self-magnetic fields are much stronger,
the formation of VC is determined mainly by the mechanism of vortex
electron structure formation, whereas the suppression of REB
transversal dynamics with the growth of external magnetic field has
a minor role. So, the first physical mechanism dominates the second
one considerably. As a consequence the critical current dependencies
in this case (see curves\,3 and 4 in Fig.~\ref{Pic:CrB}a)
demonstrate the immediate increase with the external magnetic field
growth ($0\leq B_0\leq B_{max}^i$).

Taking into account that the external magnetic field $B_0\geq B_{min}^i$ is
strong enough to neglect the influence of the self-magnetic fields,
the value $B_{min}^i$  may be easily estimated analytically.
Let the REB with current $I_{cr}$ has the radius $R_b$
at the injection plane and the character radius $R_{VC}$ --- at the
VC region ($R_{VC}>R_b$ as a result of Coulomb's repulsion of
electrons of beam; the value of $R_{VC}$ is taken from the simulation). Moving in
the system between the points with radii $R_b$ and $R_{VC}$ in the
presence of the external magnetic field, the electrons acquire the angular momentum. This momentum is proportional
to the difference of induction fluxes across the cross-sections of
the REB at the points with radii $R_b$ and $R_{VC}$,
respectively~\cite{Tsimring:2007_Beams}:
\begin{equation}
R_{VC}^2\frac{d\theta}{dt}=\frac{\pi\eta B_0}{2\pi\gamma_0}\left(R_{VC}^2-
R_b^2 \right), \label{eq:Busch}
\end{equation}
where $d\theta/dt$ is the azimuthal velocity of electrons. The motion of
electrons of REB is determined by the action of centrifugal force
$F_c=\gamma_0 m_er\left(d\theta/dt\right)^2$, Coulomb's repulsion
force $F_k=-eE_r$ and Lorentz force ${F_L=-er(d\theta/dt)B}$ (here $e$
and $m_e$ are the charge and the mass of the electron, respectively, $r$ is
the radial coordinate of electron, $E_r$ is the radial component of
the space charge field intensity). Then, one can write the motion
equation for the boundary electron of the beam, taking into
account the above, the relation (\ref{eq:Busch}) and equation
${d^2r}/{dt^2}=(2{\eta V_0}/{\gamma_0}){d^2r}/{dz^2}$:
\begin{equation}
\frac{d^2r}{dz^2}+\frac{\eta
B_0^2}{8V_0\gamma_0}R_{VC}\left[1-\left(\frac{R_b}{R_{VC}}\right)^4\right]
-\frac{I\sqrt{\gamma_0}}{4\pi\varepsilon_0\sqrt{2\eta}V_0^{3/2}R_{VC}}=0,
\label{eq:Traj}
\end{equation}
where $V_0$ is  the accelerating voltage.

Eq.~(\ref{eq:Traj}) implies that there is such external
magnetic field, $B_{min}^i$, for which REB keeps the constant radius
in the system. Actually, if we put $d^2r/dz^2=0$ in Eq.~(\ref{eq:Traj}) (it means the lack of the acceleration in the radial
direction), we obtain the quadratic equation for $B_0$. The solution of
this equation gives the required value of the external magnetic field
$B_{min}^i$ when the dependency $I_{cr}(B_0)$ saturates (see
Fig.~\ref{Pic:CrB}):
\begin{equation}
B_{min}^i=R_{VC}\sqrt{\frac{\sqrt{2}I\gamma_0^{3/2}}{\pi\varepsilon_0\eta^{3/2}\sqrt{V_0}(R_{VC}^4-R_b^4)}}.
\label{eq:Bmath}
\end{equation}

\begin{figure}[h]
\begin{center}
\includegraphics[width=8.2cm]{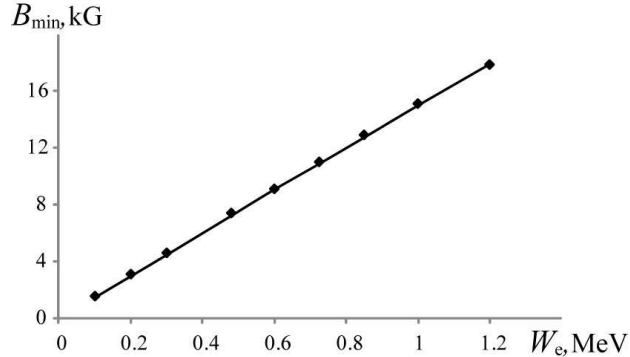}
\end{center}
\caption{The comparison of the external magnetic field value $B_{min}^i$ defined according to Eq.~(\ref{eq:Bmath}) (solid curve) with the value $B_{min}^i$ calculated with the help of 3D numerical simulation (dots) for annular REB \label{Pic:Bmin(We)}}
\end{figure}

Fig.~\ref{Pic:Bmin(We)} demonstrates analytical curve
(\ref{eq:Bmath}) (solid line) and the numerically calculated values
(points) of the external magnetic field $B^i_{min}$ for the different
beam energies $W_e$. One can see that relation~(\ref{eq:Bmath})
describes the results of the numerical simulation correctly and gives the
accurate values of the magnetic field $B^i_{min}$ being close to
the corresponding numerical results. The value $B^i_{min}$ increases
with the growth of the beam energy, $W_e$, due to the following reason. We
would remind that $B_{min}$ is such external magnetic field value
for which the transversal dynamics of electron beam is completely
suppressed in the system. Obviously, the electron inertia increases with
the growth of the beam energy, $W_e$, due to the electron velocities
increase and, as a consequence, the relativistic growth of the electron mass
is observed. As a result, the greater external magnetic field value is
required for the electrons in the system to be focused (the limitation of
the beam transversal dynamics), therefore, $B_{min}$ increases
monotonously with the growth of the REB energy. The behavior of the dependency
$B_{min}(W_e)$ on the energy $W_e$ is qualitatively similar to the
weakly and strong relativistic cases (see Fig.~\ref{Pic:Bmin(We)}).

\section{Effect of the electron beam thickness on the virtual cathode formation conditions}
\label{sct:thickness} In the previous sections we have analyzed
the dynamics of a thin annular REB with a fixed thickness, $d$ (see
Section~\ref{sct:model}). Let us consider here the results of
numerical simulation of the influence of the beam thickness, $d$, on
the critical current of REB as well as on the dynamics of the vortex
structure in the system.

\begin{figure}[h]
  \includegraphics[width=8.2cm]{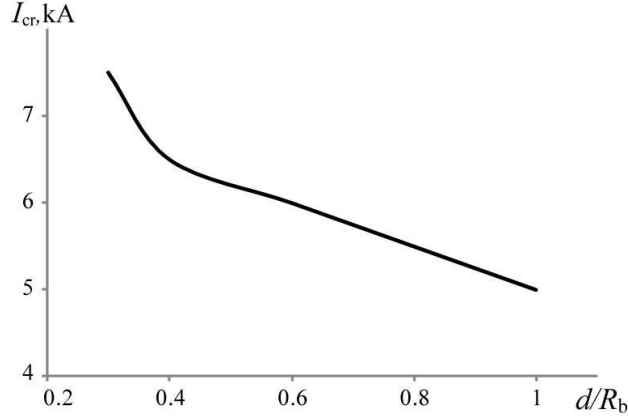}
\caption{Dependency of the REB critical current on the normalized beam thickness $d/R_b$ for
the external magnetic field induction value $B_0=3$\,kGs
and the electron beam initial energy value $W_e=480$\,keV}
\label{Pic:CrD}
\end{figure}

Fig.~\ref{Pic:CrD} illustrates the dependency of the REB critical
current on the beam thickness $d$. One can see that the critical
current value decreases monotonously with the growth of $d$. More
homogeneous filling of the drift tube by the electron beam with the
increasing of the beam thickness results in greater potential
sagging in the beam drift space. As a consequence, the beam critical
current decreases with the growth of the beam thickness that agrees
qualitatively with the analytical dependency of the critical current
value on the beam thickness in the case of 1D motion of electron
beam (\ref{I_0_B=INFTY}) \cite{Bogdankevich:1971_Engl}.

\begin{figure}[h]
  \includegraphics[width=8.2cm]{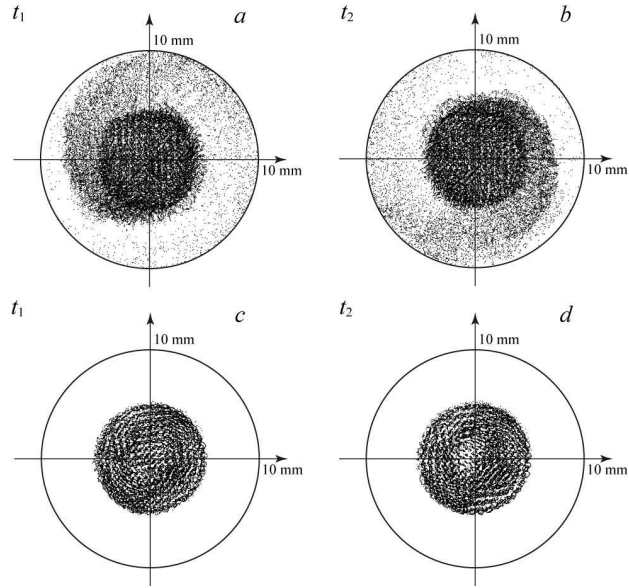}
    \caption{Projections of the instantaneous positions of charged particles of the solid electron beam at the transverse plane $(r,\theta)$ for $z_s=9.5$\,mm at the consecutive time points $t_1$ and $t_2$ ($t_2-t_1=0.2$\,ns) for $B_0=3$\,kGs (upper figures) and $B_0=30$\,kGs (lower figures); $I=7.5$\,kA, $W_e=480$\,keV. Only particles behind the projection plane ($z>z_s$) are shown in Figures }
    \label{Pic:REB_solid}
\end{figure}


Fig.~\ref{Pic:REB_solid} shows the typical projections of the
instantaneous positions of charged particles of the solid REB at the
transverse plane $(r,\theta)$ for the beam current being greater the
critical value and the different characteristic values of the
external magnetic field, $B_0$. Analyzing these figures, one can see
that the vortex structure in the solid REB is formed (see
Fig.~\ref{Pic:REB_solid}\textit{a,b}) similarly to the case of the
annular electron beam in the weak external magnetic fields (see
Section~\ref{sct:PhysProc}; compare
Fig.~\ref{Pic:REB_solid}\textit{a,b} and Fig.~\ref{Pic:REB1}). Let
us note that the obtained results for the solid REB agree
qualitatively with the experimental investigation of the intense REB
carried out in the Ref.\cite{1984:REB_vortex} where the tracings of
witness place damage patterns (see Fig.~6 (left panel) in
Ref.~\cite{1984:REB_vortex}) demonstrate the presence of the vortex
structure in the solid beam in the case of weak external focusing
magnetic field. The same result was obtained for different beam
thicknesses. In particular, with increase of external magnetic field
we observe the suppression of the vortex pattern formation (see
Fig.~\ref{Pic:REB_solid}\textit{c,d}) which is also in good
agreement with an experiment (see Fig.~6 (right panel) in
Ref.~\cite{1984:REB_vortex}).


\section{Conclusion}
\label{sct:concl}

Summarizing the obtained results, it is obvious that the VC
formation and electron structure dynamics in the REB differ
considerably in the cases of the weakly and ultra relativistic
beams. We shown that the nonlinear dynamics of the REB in the weak
external magnetic fields is defined by the self-magnetic field of
REB which leads to the azimuthal instability and the vortex electron
structure formation. Such behavior reduces the value of the REB
critical current. The growth of the external magnetic field causes
the suppression of the azimuthal instability and, consequently, the
critical current increases. In the case of the strong external
magnetic field the critical current value decreases and saturates
similarly to the weakly relativistic case~\cite{Kurkin:2009_PPR,
Hramov:2008_Critical_Current}. The beam thickness does not
effect considerably on the process of the vortex structure
formation, however, the growth of the beam thickness leads to the
decrease of the beam critical current.

\begin{acknowledgments}
We thank Prof. I.I.~Magda for the fruitful discussion of the obtained
results. This work has been supported by RFBR (12-02-00345,
12-02-90022), Federal special-purpose programme
``Scientific and educational personnel of innovation Russia''
(2009–-2013) and ``Dynasty'' Foundation.
\end{acknowledgments}


\begin{thebibliography}{41}
\expandafter\ifx\csname natexlab\endcsname\relax\def\natexlab#1{#1}\fi
\expandafter\ifx\csname bibnamefont\endcsname\relax
  \def\bibnamefont#1{#1}\fi
\expandafter\ifx\csname bibfnamefont\endcsname\relax
  \def\bibfnamefont#1{#1}\fi
\expandafter\ifx\csname citenamefont\endcsname\relax
  \def\citenamefont#1{#1}\fi
\expandafter\ifx\csname url\endcsname\relax
  \def\url#1{\texttt{#1}}\fi
\expandafter\ifx\csname urlprefix\endcsname\relax\def\urlprefix{URL }\fi
\providecommand{\bibinfo}[2]{#2}
\providecommand{\eprint}[2][]{\url{#2}}

\bibitem[{\citenamefont{Sullivan et~al.}(1987)\citenamefont{Sullivan, Walsh,
  and Coutsias}}]{Sullivan:1987_VCO_Book}
\bibinfo{author}{\bibfnamefont{D.~J.} \bibnamefont{Sullivan}},
  \bibinfo{author}{\bibfnamefont{J.~E.} \bibnamefont{Walsh}}, \bibnamefont{and}
  \bibinfo{author}{\bibfnamefont{E.~A.} \bibnamefont{Coutsias}},
  \emph{\bibinfo{title}{Virtual cathode oscillator (vircator) theory}},
  vol.~\bibinfo{volume}{13} of \emph{\bibinfo{series}{High Power Microwave
  Sources}} (\bibinfo{publisher}{Artech House Microwave Library},
  \bibinfo{year}{1987}), \bibinfo{edition}{granatstein, v.l. and alexeff, i.}
  ed.

\bibitem[{\citenamefont{Alyokhin et~al.}(1994)\citenamefont{Alyokhin, Dubinov,
  Selemir, Shamro, Stepanov, and Vatrunin}}]{Alyokhin:1994}
\bibinfo{author}{\bibfnamefont{V.~D.} \bibnamefont{Alyokhin}},
  \bibinfo{author}{\bibfnamefont{A.~E.} \bibnamefont{Dubinov}},
  \bibinfo{author}{\bibfnamefont{V.~D.} \bibnamefont{Selemir}},
  \bibinfo{author}{\bibfnamefont{O.~A.} \bibnamefont{Shamro}},
  \bibinfo{author}{\bibfnamefont{N.~V.} \bibnamefont{Stepanov}},
  \bibnamefont{and} \bibinfo{author}{\bibfnamefont{V.~E.}
  \bibnamefont{Vatrunin}}, \bibinfo{journal}{IEEE Trans. Plasma Sci.}
  \textbf{\bibinfo{volume}{22}}, \bibinfo{pages}{954} (\bibinfo{year}{1994}).

\bibitem[{\citenamefont{Dubinov and Selemir}(2002)}]{Dubinov:2002_ReviewEngl}
\bibinfo{author}{\bibfnamefont{A.~E.} \bibnamefont{Dubinov}} \bibnamefont{and}
  \bibinfo{author}{\bibfnamefont{V.~D.} \bibnamefont{Selemir}},
  \bibinfo{journal}{Journal of Communications Technology and Electronics}
  \textbf{\bibinfo{volume}{47}}, \bibinfo{pages}{575} (\bibinfo{year}{2002}).

\bibitem[{\citenamefont{Kalinin et~al.}(2005)\citenamefont{Kalinin,
  Koronovskii, Hramov, Egorov, and Filatov}}]{Kalinin:2005_PhysPlazma_ENG}
\bibinfo{author}{\bibfnamefont{Y.}~\bibnamefont{Kalinin}},
  \bibinfo{author}{\bibfnamefont{A.~A.} \bibnamefont{Koronovskii}},
  \bibinfo{author}{\bibfnamefont{A.~E.} \bibnamefont{Hramov}},
  \bibinfo{author}{\bibfnamefont{E.~N.} \bibnamefont{Egorov}},
  \bibnamefont{and} \bibinfo{author}{\bibfnamefont{R.~A.}
  \bibnamefont{Filatov}}, \bibinfo{journal}{Plasma Phys. Reports}
  \textbf{\bibinfo{volume}{31}}, \bibinfo{pages}{938} (\bibinfo{year}{2005}).

\bibitem[{\citenamefont{Benford et~al.}(2007)\citenamefont{Benford, Swegle, and
  Schamiloglu}}]{Benford:2007_book}
\bibinfo{author}{\bibfnamefont{J.}~\bibnamefont{Benford}},
  \bibinfo{author}{\bibfnamefont{J.~A.} \bibnamefont{Swegle}},
  \bibnamefont{and}
  \bibinfo{author}{\bibfnamefont{E.}~\bibnamefont{Schamiloglu}},
  \emph{\bibinfo{title}{High Power Microwaves}} (\bibinfo{publisher}{CRC Press,
  Taylor and Francis}, \bibinfo{year}{2007}).

\bibitem[{\citenamefont{Biswas}(2009)}]{biswas:063104}
\bibinfo{author}{\bibfnamefont{D.}~\bibnamefont{Biswas}},
  \bibinfo{journal}{Physics of Plasmas} \textbf{\bibinfo{volume}{16}},
  \bibinfo{eid}{063104} (pages~\bibinfo{numpages}{6}) (\bibinfo{year}{2009}).

\bibitem[{\citenamefont{Filatov et~al.}(2009)\citenamefont{Filatov, Hramov,
  Bliokh, Koronovskii, and Felsteiner}}]{filatov:033106}
\bibinfo{author}{\bibfnamefont{R.~A.} \bibnamefont{Filatov}},
  \bibinfo{author}{\bibfnamefont{A.~E.} \bibnamefont{Hramov}},
  \bibinfo{author}{\bibfnamefont{Y.~P.} \bibnamefont{Bliokh}},
  \bibinfo{author}{\bibfnamefont{A.~A.} \bibnamefont{Koronovskii}},
  \bibnamefont{and}
  \bibinfo{author}{\bibfnamefont{J.}~\bibnamefont{Felsteiner}},
  \bibinfo{journal}{Physics of Plasmas} \textbf{\bibinfo{volume}{16}},
  \bibinfo{pages}{033106} (\bibinfo{year}{2009}).

\bibitem[{\citenamefont{Ender et~al.}(2011)\citenamefont{Ender, Kuznetsov, and
  Schamel}}]{ender:033502}
\bibinfo{author}{\bibfnamefont{A.~Y.} \bibnamefont{Ender}},
  \bibinfo{author}{\bibfnamefont{V.~I.} \bibnamefont{Kuznetsov}},
  \bibnamefont{and} \bibinfo{author}{\bibfnamefont{H.}~\bibnamefont{Schamel}},
  \bibinfo{journal}{Physics of Plasmas} \textbf{\bibinfo{volume}{18}},
  \bibinfo{eid}{033502} (pages~\bibinfo{numpages}{11}) (\bibinfo{year}{2011}).

\bibitem[{\citenamefont{Singh and Shashank}(2011)}]{Singh:2011_VC}
\bibinfo{author}{\bibfnamefont{G.}~\bibnamefont{Singh}} \bibnamefont{and}
  \bibinfo{author}{\bibfnamefont{C.}~\bibnamefont{Shashank}},
  \bibinfo{journal}{Physics of Plasmas} \textbf{\bibinfo{volume}{18}},
  \bibinfo{pages}{063104} (\bibinfo{year}{2011}).

\bibitem[{\citenamefont{Hramov et~al.}(2010)\citenamefont{Hramov, Koronovskii,
  and Kurkin}}]{Hramov:2010_MPFS}
\bibinfo{author}{\bibfnamefont{A.~E.} \bibnamefont{Hramov}},
  \bibinfo{author}{\bibfnamefont{A.~A.} \bibnamefont{Koronovskii}},
  \bibnamefont{and} \bibinfo{author}{\bibfnamefont{S.~A.}
  \bibnamefont{Kurkin}}, \bibinfo{journal}{Phys. Lett. A}
  \textbf{\bibinfo{volume}{374}}, \bibinfo{pages}{3057} (\bibinfo{year}{2010}).

\bibitem[{\citenamefont{Trubetskov et~al.}(1996)\citenamefont{Trubetskov,
  Mchedlova, Anfinogentov, Ponomarenko, and Ryskin}}]{Trubetskov:1996_CHAOS}
\bibinfo{author}{\bibfnamefont{D.~I.} \bibnamefont{Trubetskov}},
  \bibinfo{author}{\bibfnamefont{E.~S.} \bibnamefont{Mchedlova}},
  \bibinfo{author}{\bibfnamefont{V.~G.} \bibnamefont{Anfinogentov}},
  \bibinfo{author}{\bibfnamefont{V.~I.} \bibnamefont{Ponomarenko}},
  \bibnamefont{and} \bibinfo{author}{\bibfnamefont{N.~M.}
  \bibnamefont{Ryskin}}, \bibinfo{journal}{Chaos} \textbf{\bibinfo{volume}{6}},
  \bibinfo{pages}{358} (\bibinfo{year}{1996}).

\bibitem[{\citenamefont{Anfinogentov and
  Hramov}(1998)}]{Anfinogentov:1998_VCO_ENGL}
\bibinfo{author}{\bibfnamefont{V.~G.} \bibnamefont{Anfinogentov}}
  \bibnamefont{and} \bibinfo{author}{\bibfnamefont{A.~E.}
  \bibnamefont{Hramov}}, \bibinfo{journal}{Radiophysics and Quantum
  Electronics} \textbf{\bibinfo{volume}{41}}, \bibinfo{pages}{1137}
  (\bibinfo{year}{1998}).

\bibitem[{\citenamefont{Koronovskii and
  Hramov}(2002)}]{Koronovskii:2002_PierceEnglish}
\bibinfo{author}{\bibfnamefont{A.~A.} \bibnamefont{Koronovskii}}
  \bibnamefont{and} \bibinfo{author}{\bibfnamefont{A.~E.}
  \bibnamefont{Hramov}}, \bibinfo{journal}{Plasma Physics Reports}
  \textbf{\bibinfo{volume}{28}}, \bibinfo{pages}{666} (\bibinfo{year}{2002}).

\bibitem[{\citenamefont{Hramov et~al.}(2011)\citenamefont{Hramov, Koronovsky,
  Kurkin, and Rempen}}]{hramov:2011_IJE}
\bibinfo{author}{\bibfnamefont{A.~E.} \bibnamefont{Hramov}},
  \bibinfo{author}{\bibfnamefont{A.~A.} \bibnamefont{Koronovsky}},
  \bibinfo{author}{\bibfnamefont{S.~A.} \bibnamefont{Kurkin}},
  \bibnamefont{and} \bibinfo{author}{\bibfnamefont{I.~S.}
  \bibnamefont{Rempen}}, \bibinfo{journal}{Int. J. Electronics}
  \textbf{\bibinfo{volume}{98}}, \bibinfo{pages}{1549} (\bibinfo{year}{2011}).

\bibitem[{\citenamefont{Mahaffey et~al.}(1977)\citenamefont{Mahaffey, Sprangle,
  Golden, and Kapetanakos}}]{Mahaffey:1977_First_VC}
\bibinfo{author}{\bibfnamefont{R.~A.} \bibnamefont{Mahaffey}},
  \bibinfo{author}{\bibfnamefont{P.~A.} \bibnamefont{Sprangle}},
  \bibinfo{author}{\bibfnamefont{J.}~\bibnamefont{Golden}}, \bibnamefont{and}
  \bibinfo{author}{\bibfnamefont{C.~A.} \bibnamefont{Kapetanakos}},
  \bibinfo{journal}{Phys.Rev.Lett} \textbf{\bibinfo{volume}{39}},
  \bibinfo{pages}{843} (\bibinfo{year}{1977}).

\bibitem[{\citenamefont{Sze et~al.}(1986)\citenamefont{Sze, Benford, and
  Harteneck}}]{Sze:1986_VC}
\bibinfo{author}{\bibfnamefont{H.}~\bibnamefont{Sze}},
  \bibinfo{author}{\bibfnamefont{J.}~\bibnamefont{Benford}}, \bibnamefont{and}
  \bibinfo{author}{\bibfnamefont{B.}~\bibnamefont{Harteneck}},
  \bibinfo{journal}{Phys. Fluids} \textbf{\bibinfo{volume}{29}},
  \bibinfo{pages}{5875} (\bibinfo{year}{1986}).

\bibitem[{\citenamefont{Burkhart et~al.}(1985)\citenamefont{Burkhart,
  Scarpetty, and Lundberg}}]{Burkhart:1985_VCO}
\bibinfo{author}{\bibfnamefont{S.~C.} \bibnamefont{Burkhart}},
  \bibinfo{author}{\bibfnamefont{R.~D.} \bibnamefont{Scarpetty}},
  \bibnamefont{and} \bibinfo{author}{\bibfnamefont{R.~L.}
  \bibnamefont{Lundberg}}, \bibinfo{journal}{J.Appl.Phys.}
  \textbf{\bibinfo{volume}{58}}, \bibinfo{pages}{28} (\bibinfo{year}{1985}).

\bibitem[{\citenamefont{Hoeberling and
  Fazio}(1992)}]{Hoeberling:1992_AdvanceVCO}
\bibinfo{author}{\bibfnamefont{R.~F.} \bibnamefont{Hoeberling}}
  \bibnamefont{and} \bibinfo{author}{\bibfnamefont{M.~V.} \bibnamefont{Fazio}},
  \bibinfo{journal}{IEEE Trans. Electromagnetic Compatibility}
  \textbf{\bibinfo{volume}{34}}, \bibinfo{pages}{252} (\bibinfo{year}{1992}).

\bibitem[{\citenamefont{Bogdankevich and
  Rukhadze}(1971)}]{Bogdankevich:1971_Engl}
\bibinfo{author}{\bibfnamefont{L.~A.} \bibnamefont{Bogdankevich}}
  \bibnamefont{and} \bibinfo{author}{\bibfnamefont{A.~A.}
  \bibnamefont{Rukhadze}}, \bibinfo{journal}{Sov. Phys. Uspekhi}
  \textbf{\bibinfo{volume}{14}}, \bibinfo{pages}{163} (\bibinfo{year}{1971}).

\bibitem[{\citenamefont{Jiang et~al.}(1995)\citenamefont{Jiang, Masugata, and
  Yatsui}}]{Jiang1995_Mech_VC}
\bibinfo{author}{\bibfnamefont{W.}~\bibnamefont{Jiang}},
  \bibinfo{author}{\bibfnamefont{K.}~\bibnamefont{Masugata}}, \bibnamefont{and}
  \bibinfo{author}{\bibfnamefont{K.}~\bibnamefont{Yatsui}},
  \bibinfo{journal}{Phys. Plasmas} \textbf{\bibinfo{volume}{2}},
  \bibinfo{pages}{982} (\bibinfo{year}{1995}).

\bibitem[{\citenamefont{Hramov}(1999)}]{Hramov:1999_ChaosVCO}
\bibinfo{author}{\bibfnamefont{A.~E.} \bibnamefont{Hramov}},
  \bibinfo{journal}{J. Communication Technology and Electronics}
  \textbf{\bibinfo{volume}{44}}, \bibinfo{pages}{551} (\bibinfo{year}{1999}).

\bibitem[{\citenamefont{Egorov et~al.}(2007)\citenamefont{Egorov, Kalinin,
  Koronovskii, Levin, and Hramov}}]{Egorov_1D:2007}
\bibinfo{author}{\bibfnamefont{E.~N.} \bibnamefont{Egorov}},
  \bibinfo{author}{\bibfnamefont{Y.}~\bibnamefont{Kalinin}},
  \bibinfo{author}{\bibfnamefont{A.~A.} \bibnamefont{Koronovskii}},
  \bibinfo{author}{\bibfnamefont{Y.}~\bibnamefont{Levin}}, \bibnamefont{and}
  \bibinfo{author}{\bibfnamefont{A.~E.} \bibnamefont{Hramov}},
  \bibinfo{journal}{J. Comm. Techn. Electron.} \textbf{\bibinfo{volume}{52}},
  \bibinfo{pages}{45} (\bibinfo{year}{2007}).

\bibitem[{\citenamefont{Lindsay et~al.}(2002)\citenamefont{Lindsay, Toh, and
  Chen}}]{Lindsay:2002_VCO}
\bibinfo{author}{\bibfnamefont{P.~A.} \bibnamefont{Lindsay}},
  \bibinfo{author}{\bibfnamefont{W.~K.} \bibnamefont{Toh}}, \bibnamefont{and}
  \bibinfo{author}{\bibfnamefont{X.}~\bibnamefont{Chen}},
  \bibinfo{journal}{IEEE Trans. Plasma Sci.} \textbf{\bibinfo{volume}{30}},
  \bibinfo{pages}{1186} (\bibinfo{year}{2002}).

\bibitem[{\citenamefont{Chen et~al.}(2004)\citenamefont{Chen, Toh, and
  Lindsay}}]{Chen:2004_VCO_Coaxial}
\bibinfo{author}{\bibfnamefont{X.}~\bibnamefont{Chen}},
  \bibinfo{author}{\bibfnamefont{W.~K.} \bibnamefont{Toh}}, \bibnamefont{and}
  \bibinfo{author}{\bibfnamefont{P.~A.} \bibnamefont{Lindsay}},
  \bibinfo{journal}{IEEE Trans. Plasma Sci.} \textbf{\bibinfo{volume}{32}},
  \bibinfo{pages}{835} (\bibinfo{year}{2004}).

\bibitem[{\citenamefont{Hramov et~al.}(2008)\citenamefont{Hramov, Koronovskii,
  Morozov, and Mushtakov}}]{Hramov:2008_Critical_Current}
\bibinfo{author}{\bibfnamefont{A.~E.} \bibnamefont{Hramov}},
  \bibinfo{author}{\bibfnamefont{A.~A.} \bibnamefont{Koronovskii}},
  \bibinfo{author}{\bibfnamefont{M.}~\bibnamefont{Morozov}}, \bibnamefont{and}
  \bibinfo{author}{\bibfnamefont{A.~V.} \bibnamefont{Mushtakov}},
  \bibinfo{journal}{Phys. Lett. A} \textbf{\bibinfo{volume}{372}},
  \bibinfo{pages}{876} (\bibinfo{year}{2008}).

\bibitem[{\citenamefont{Balakirev et~al.}(1997)\citenamefont{Balakirev,
  Gorban', Magda, Novikov, and Onishchenko}}]{Magda:1997}
\bibinfo{author}{\bibfnamefont{V.~A.} \bibnamefont{Balakirev}},
  \bibinfo{author}{\bibfnamefont{A.~M.} \bibnamefont{Gorban'}},
  \bibinfo{author}{\bibfnamefont{I.~I.} \bibnamefont{Magda}},
  \bibinfo{author}{\bibfnamefont{V.~E.} \bibnamefont{Novikov}},
  \bibnamefont{and} \bibinfo{author}{\bibfnamefont{I.~N.}
  \bibnamefont{Onishchenko}}, \bibinfo{journal}{Plasma Physics Report}
  \textbf{\bibinfo{volume}{23}}, \bibinfo{pages}{323} (\bibinfo{year}{1997}).

\bibitem[{\citenamefont{Tsimring}(2007)}]{Tsimring:2007_Beams}
\bibinfo{author}{\bibfnamefont{S.~E.} \bibnamefont{Tsimring}},
  \emph{\bibinfo{title}{Electron beams and microwave vacuum electronics}}
  (\bibinfo{publisher}{John Wiley and Sons, Inc., Hoboken, New Jersey},
  \bibinfo{year}{2007}).

\bibitem[{\citenamefont{Birdsall and Langdon}(1985)}]{Birdsall:1985_PlasmaBook}
\bibinfo{author}{\bibfnamefont{C.~K.} \bibnamefont{Birdsall}} \bibnamefont{and}
  \bibinfo{author}{\bibfnamefont{A.~B.} \bibnamefont{Langdon}},
  \emph{\bibinfo{title}{Plasma physics, via computer simulation}}
  (\bibinfo{publisher}{NY: McGraw-Hill}, \bibinfo{year}{1985}).

\bibitem[{\citenamefont{Anderson et~al.}(1999)\citenamefont{Anderson, Mondelli,
  Levush, Verboncoeur, and Birdsall}}]{Anderson:1999_EM_Simulations}
\bibinfo{author}{\bibfnamefont{T.~M.} \bibnamefont{Anderson}},
  \bibinfo{author}{\bibfnamefont{A.~A.} \bibnamefont{Mondelli}},
  \bibinfo{author}{\bibfnamefont{B.}~\bibnamefont{Levush}},
  \bibinfo{author}{\bibfnamefont{J.~P.} \bibnamefont{Verboncoeur}},
  \bibnamefont{and} \bibinfo{author}{\bibfnamefont{C.~K.}
  \bibnamefont{Birdsall}}, \bibinfo{journal}{Proceedings IEEE}
  \textbf{\bibinfo{volume}{87}}, \bibinfo{pages}{804} (\bibinfo{year}{1999}).

\bibitem[{\citenamefont{Egorov and Hramov}(2006)}]{Egorov:2006_2.5DVCO_Engl}
\bibinfo{author}{\bibfnamefont{E.~N.} \bibnamefont{Egorov}} \bibnamefont{and}
  \bibinfo{author}{\bibfnamefont{A.~E.} \bibnamefont{Hramov}},
  \bibinfo{journal}{Plasma Physics Reports} \textbf{\bibinfo{volume}{32}},
  \bibinfo{pages}{683} (\bibinfo{year}{2006}).

\bibitem[{\citenamefont{Boris and Lee}(1969)}]{Boris:1969}
\bibinfo{author}{\bibfnamefont{J.~P.} \bibnamefont{Boris}} \bibnamefont{and}
  \bibinfo{author}{\bibfnamefont{R.}~\bibnamefont{Lee}},
  \bibinfo{journal}{Commun. Math. Phys.} \textbf{\bibinfo{volume}{12}},
  \bibinfo{pages}{131} (\bibinfo{year}{1969}).

\bibitem[{\citenamefont{Rouch}(1976)}]{Rouch:1976_FluidNumericalBook}
\bibinfo{author}{\bibfnamefont{P.~J.} \bibnamefont{Rouch}},
  \emph{\bibinfo{title}{Computational fluid dynamics}}
  (\bibinfo{publisher}{Hermosa publishers, Albuquerque}, \bibinfo{year}{1976}).

\bibitem[{\citenamefont{Granatstein and
  Alexeeff}(1987)}]{Granatstein:1987_Book}
\bibinfo{author}{\bibfnamefont{V.~L.} \bibnamefont{Granatstein}}
  \bibnamefont{and} \bibinfo{author}{\bibfnamefont{I.}~\bibnamefont{Alexeeff}},
  \emph{\bibinfo{title}{High Power Microwave Sources}}
  (\bibinfo{publisher}{Artech House Microwave Library}, \bibinfo{year}{1987}).

\bibitem[{\citenamefont{Kurkin et~al.}(2009{\natexlab{a}})\citenamefont{Kurkin,
  Hramov, and Koronovskii}}]{Kurkin:2009_PPR}
\bibinfo{author}{\bibfnamefont{S.~A.} \bibnamefont{Kurkin}},
  \bibinfo{author}{\bibfnamefont{A.~E.} \bibnamefont{Hramov}},
  \bibnamefont{and} \bibinfo{author}{\bibfnamefont{A.~A.}
  \bibnamefont{Koronovskii}}, \bibinfo{journal}{Plasma Phys. Report}
  \textbf{\bibinfo{volume}{35}}, \bibinfo{pages}{628}
  (\bibinfo{year}{2009}{\natexlab{a}}).

\bibitem[{\citenamefont{Luginsland et~al.}(1998)\citenamefont{Luginsland,
  McGee, and Lau}}]{700866}
\bibinfo{author}{\bibfnamefont{J.~W.} \bibnamefont{Luginsland}},
  \bibinfo{author}{\bibfnamefont{S.}~\bibnamefont{McGee}}, \bibnamefont{and}
  \bibinfo{author}{\bibfnamefont{Y.~Y.} \bibnamefont{Lau}},
  \bibinfo{journal}{Plasma Science, IEEE Transactions on}
  \textbf{\bibinfo{volume}{26}}, \bibinfo{pages}{901} (\bibinfo{year}{1998}),
  ISSN \bibinfo{issn}{0093-3813}.

\bibitem[{\citenamefont{Biswas and Kumar}(2006)}]{biswas:073101}
\bibinfo{author}{\bibfnamefont{D.}~\bibnamefont{Biswas}} \bibnamefont{and}
  \bibinfo{author}{\bibfnamefont{R.}~\bibnamefont{Kumar}},
  \bibinfo{journal}{Physics of Plasmas} \textbf{\bibinfo{volume}{13}},
  \bibinfo{eid}{073101} (pages~\bibinfo{numpages}{6}) (\bibinfo{year}{2006}).

\bibitem[{\citenamefont{Davidson}(1974)}]{Davidson:1974}
\bibinfo{author}{\bibfnamefont{R.~C.} \bibnamefont{Davidson}},
  \emph{\bibinfo{title}{Theory of Nonneutral Plasmas}}
  (\bibinfo{publisher}{W.A. Benjamin Inc., Advanced book program},
  \bibinfo{year}{1974}).

\bibitem[{\citenamefont{Levy}(1965)}]{levy:1288}
\bibinfo{author}{\bibfnamefont{R.~H.} \bibnamefont{Levy}},
  \bibinfo{journal}{Physics of Fluids} \textbf{\bibinfo{volume}{8}},
  \bibinfo{pages}{1288} (\bibinfo{year}{1965}).

\bibitem[{\citenamefont{Peratt and Snell}(1985)}]{PhysRevLett.54.1167}
\bibinfo{author}{\bibfnamefont{A.~L.} \bibnamefont{Peratt}} \bibnamefont{and}
  \bibinfo{author}{\bibfnamefont{C.~M.} \bibnamefont{Snell}},
  \bibinfo{journal}{Phys. Rev. Lett.} \textbf{\bibinfo{volume}{54}},
  \bibinfo{pages}{1167} (\bibinfo{year}{1985}).

\bibitem[{\citenamefont{Kurkin et~al.}(2009{\natexlab{b}})\citenamefont{Kurkin,
  Koronovskii, and Hramov}}]{Kurkin:2009_JTF_VCMagnetic_Engl}
\bibinfo{author}{\bibfnamefont{S.~A.} \bibnamefont{Kurkin}},
  \bibinfo{author}{\bibfnamefont{A.~A.} \bibnamefont{Koronovskii}},
  \bibnamefont{and} \bibinfo{author}{\bibfnamefont{A.~E.}
  \bibnamefont{Hramov}}, \bibinfo{journal}{Technical Physics}
  \textbf{\bibinfo{volume}{54}}, \bibinfo{pages}{1520}
  (\bibinfo{year}{2009}{\natexlab{b}}).

\bibitem[{\citenamefont{O'Shea et~al.}(1984)\citenamefont{O'Shea, Welsh,
  Destler, and Striffler}}]{1984:REB_vortex}
\bibinfo{author}{\bibfnamefont{P.~G.} \bibnamefont{O'Shea}},
  \bibinfo{author}{\bibfnamefont{D.}~\bibnamefont{Welsh}},
  \bibinfo{author}{\bibfnamefont{W.~W.} \bibnamefont{Destler}},
  \bibnamefont{and} \bibinfo{author}{\bibfnamefont{C.~D.}
  \bibnamefont{Striffler}}, \bibinfo{journal}{J. Appl. Phys.}
  \textbf{\bibinfo{volume}{55}}, \bibinfo{pages}{3934} (\bibinfo{year}{1984}).

\end{thebibliography}

\end{document}